# High School Science Profile Predicts Adults' Views on the Future of AI and STS

Gyeonggeon "Boaz" Lee*

*Natural Sciences and Science Education Department, National Institute of Education, Nanyang Technological University, 1 Nanyang Walk, Singapore 637616, Singapore*

*corresponding author: gyeonggeon.lee@nie.edu.sg

## Abstract

This study investigates the long-term influence of high school science education on adults' engagement with artificial intelligence (AI) and their views on science–technology–society (STS) issues. Drawing on longitudinal data from the Korea Employment Education Panel (KEEP) II ($n = 2{,}348$), which tracked general high school students from 2016 to 2023, we applied structural equation modeling (SEM) to examine how science interest and achievement in adolescence predict AI use and perceptions in adulthood. Results indicate that high school science achievement, but not science interest, directly predicted AI use at age 24. AI use significantly influenced both positive and negative perceptions of AI, which in turn shaped sophisticated perspectives on STS domains—human–AI relationship, quality of life, and science and technology monopolization. Indirect effects suggest that high school science interest can influence adult perceptions of AI and STS views, mediated by science achievement and AI use. These findings provide rare empirical evidence linking secondary science education to adult engagement with AI.

## Introduction and Backgrounds

In every generation, science education has been an ever-present endeavor for the future, bridging the two for students who will live out a complex world (Sjöström & Eilks, 2018). In the current status of global society, artificial intelligence (AI) is deemed to be the most powerful factor that urges science educators to rethink and renovate teaching and learning, so that students can thrive in the unforeseen world that is facing economic, societal, and environmental challenges (UNESCO, 2021; 2023; 2024). Since the general public encountered ChatGPT and expressed their mixed reactions in 2022, science educators realized that science education is accountable for teaching about AI, to guide students for responsible and ethical use of AI (Erduran, 2023). While the cutting-edge technology can hugely influence society, if proper education to harness AI for good is not done, it may lead us into uncertainty that may endanger individual and societal well-being. For example, there are concerns about AI for its potential replacement of human workforce (UNESCO, 2021), the echo chamber effect in social media that may decrease the quality of life (Höttecke and Allchin, 2020), and misinformation or even disinformation that may benefit certain interest groups (Osborne & Pimentel, 2023), which are some aspects of the science-technology-society (STS) that are revisited in the era of AI.

However, despite the global consensus that we need to teach students about AI, there seems to be almost no empirical evidence that school period learner profile can shape AI use and perceptions in adulthood. Therefore, the researcher of this study suggests that it is necessary to investigate high school science education can lead to different AI engagement in adulthood.

In this study, we report timely evidence on the predictive factors surveyed during the school period for adults' AI use and perception. By utilizing panel data, we show that there are longitudinal predictors of AI use in the science education context.

*Research questions*

The research questions (RQs) of this study are as follows:
- RQ1. How does high school science profile predict adults' AI use?
- RQ2. How AI use experience influence perceptions of AI and views on STS?
- RQ3. How does high school science profile predict adults' perceptions of AI and views on STS, mediated by AI use?

**Methods**

*Data Preparation*

We used the Korea Employment Education Panel (KEEP) II data. The KEEP II started in 2016 with 10,558 cohorts of $11^{th}$ graders in South Korea. The panel study random-sampled high school students in Korea and has been tracking them annually.

We combined the datasets from 2016 (1st wave), 2017 (2nd wave) and 2023 (7th wave) using cohort id. We only used cohort data from general high school (not specialized high schools such as foreign language high school). We cleaned the dataset by listwise deletion if there is any missing value in the selected variables. Consequently, we prepared a dataset with the number of observations = 2,348 (male = 1,079; female = 1,269).

*Variable selection*

The list of latent and observed variables is provided in **Table 1**.

**Table 1**. Variables used in this study (○: Latent variable, □: Observed variable)

| Variable | Code | Measurement variable (scale) | Mean (sd) | Reliability (Cronbach's $\alpha$) |
|---|---|---|---|---|
| Period: High School | | | | |
| ○ Science interest | HSI1<br>HSI2<br>HSI3 | □ Science is fun<br>□ Do science well<br>□ Like science<br>(strongly disagree: 1 – strongly agree: 5) | 3.32 (1.23)<br>2.76 (1.08)<br>3.17 (1.22) | .92 |
| □ Science achievement | HAS | Stanine grade of science subject in the first semester of 12th year (2017)<br>(converted to 1: low achieving - 9: high achieving) | 5.85 (1.87) | - |
| Period: Adult (24 years old) | | | | |
| ○ AI use | AAIU1<br>AAIU2<br>AAIU3<br>AAIU4 | □ Translator service<br>□ Recommender service<br>□ Facial/image recognition service<br>□ Speech recognition service<br>(never: 1 – frequently 4) | 2.92 (.87)<br>3.27 (.89)<br>2.42 (1.03)<br>2.48 (.95) | .69 |
| ○ AI positive perception<br><br>("In 10 years, AI will enable …") | AAIP1<br>AAIP2<br>AAIP3<br>AAIP4 | □ personalized learning<br>□ efficient working<br>□ new job opportunities<br>□ remedying loneliness<br>(strongly disagree: 1 – strongly agree: 5) | 4.56 (.89)<br>4.68 (.96)<br>4.35 (1.15)<br>3.58 (1.27) | .7 |
| ○ AI negative perception<br><br>("In 10 years, due to AI, …") | AAIN1<br>AAIN2<br>AAIN3<br>AAIN4 | □ Jobs will diminish<br>□ There will be data leakage and privacy breach<br>□ Human will not make important decision making<br>□ human-human solidarity will be weakened<br>(strongly disagree: 1 – strongly agree: 5) | 4.46 (.98)<br>4.52 (1.07)<br>3.81 (1.25)<br>3.89 (1.2) | .76 |
| □ STS human and AI | STSH | "Humans have abilities that AI or robot cannot replace"<br>(strongly disagree: 1 – strongly agree: 5) | 4.65 (.99) | - |
| □ STS life quality | STSL | "The discovery of new science and technology will improve the quality of future human lives"<br>(strongly disagree: 1 – strongly agree: 5) | 4.65 (.87) | - |
| □ STS monopoly | STSM | "The benefits of new science and technology will be monopolized by few people"<br>(strongly disagree: 1 – strongly agree: 5) | 4.12 (1.08) | - |

*Statistical analysis*

We conducted structural equation modeling (SEM) analysis. Stata 16 was used for descriptive statistics and AMOS 21 was used for SEM.

*Structural equation model building*

The SEM built in this study is presented in **Figure 1-(a)**.

- *High school period*: In KEEP II, science interest was measured when the cohorts were 11$^{th}$ graders. Also, science achievement was reported in a Stanine grade for 12$^{th}$ year. Thus, it was hypothesized that science interest had affected science achievement.

Since there were year gaps between the high school science profile (2016-2017) and the adult period that reported the cohorts' views on AI and STS (2023), we did not hypothesize that the former had directly affected the latter. Rather, we hypothesized that high school science interest and achievement can affect AI use in adulthood.

- *Adult period*: The 7$^{th}$ wave of the KEEP II was administered in 2023, after the release of ChatGPT. The wave added 15 particular items that ask about the respondents' AI use, AI positive perception, AI negative perception, and views on STS to the usual items, respond to the impact of AI. Since both the positive and negative perceptions of AI ask cohorts' views on AI imagining "in 10 years" (see **Table 1**), rather than the current status, we hypothesized that AI use experience will impact their AI perceptions. Also, we hypothesized that AI use, positive perception, and negative perception will affect the three independent STS viewpoints – i.e., views on human and AI, life quality, and monopoly.

We did not merge positive and negative perceptions of AI, and the variables on STS, respectively, as the reliability was low which shows no evidence of unidimensionality.

*Model modification*

We referred to modification indices during model-fitting stage to revise the model. Consequently, we added covariance paths between positive and negative AI perceptions, and between STS views on human and AI and life quality. We also put two covariances between observed variables in measurement models (AAIN3 ↔ AAIN4; AAIU3 ↔ AAIU4).

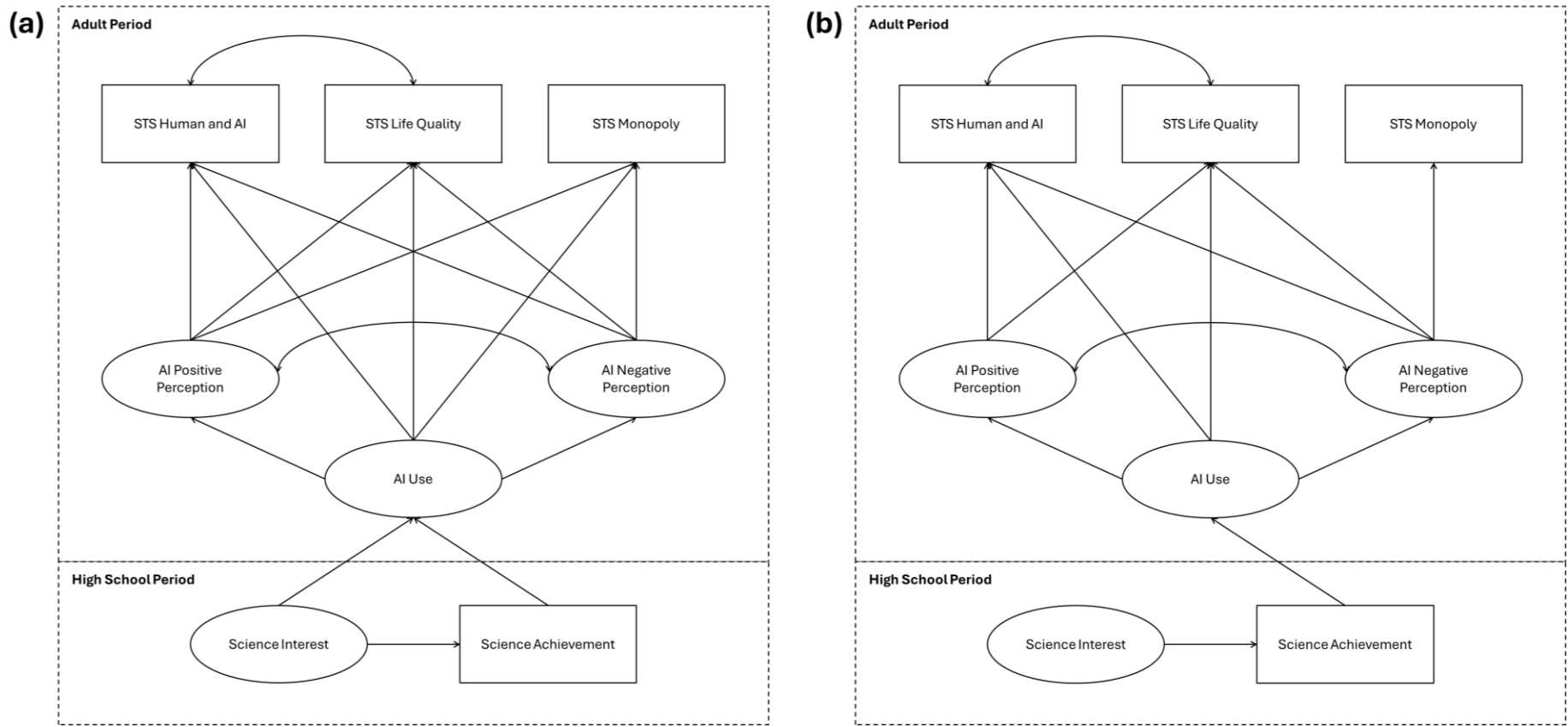

**Figure 1**. Full research model (a) and significant path model (b)

**Results**

*Descriptive statistics*

Descriptive statistics and reliability of the variables used in this study are presented in **Table 1**. Pairwise correlations of variables are presented in **Table 2**.

*Model goodness of fit*

The goodness of fit for the measurement model and the research model is given in **Table 3**. In both models, $\chi^2/df$ was larger than 2 ($p < .05$), due to the large sample size. CFI, TLI, and NFI were above .9, RMSEA was slightly larger than .05, and SRMR was less than .08, indicating a good fit (Hu & Bentler, 1998; Kline, 2015).

*Path coefficients*

The standardized path coefficients in the fitted model are presented in **Table 4**. **Figure 1-(b)** shows the model with only significant paths ($p < .05$). During high school period, as expected, science interest positively affected science achievement ($\beta = .349$, $p < .001$). However, between high school period and adult period, only science achievement showed a significant prediction path toward AI use ($\beta = .18$, $p < .001$), but science interest ($p > .05$).

AI use showed significantly positive paths both on AI positive perception ($\beta = .412$, $p < .001$) and AI negative perception ($\beta = .295$, $p < .001$). Positive and negative perceptions of AI showed a significant correlation ($\beta = .297$, $p < .001$).

STS human and AI was positively affected by AI use ($\beta = .08$, $p < .01$), and positive ($\beta = .225$, $p < .001$) and negative AI perceptions ($\beta = .134$, $p < .001$). STS life quality was also positively affected by AI use ($\beta = .118$, $p < .001$), and positive ($\beta = .472$, $p < .001$) and negative AI perceptions ($\beta = .122$, $p < .05$). However, STS monopoly was significantly influenced only by AI negative perception ($\beta = .474$, $p < .001$), not AI positive perception or AI use ($p > .05$).

**Table 2**. Goodness of fit of measurement and research models

| | HSI1 | HSI2 | HSI3 | HAS | AAIU1 | AAIU2 | AAIU3 | AAIU4 | AAIP1 | AAIP2 | AAIP3 | AAIP4 | AAIN1 | AAIN2 | AAIN3 | AAIN4 | STSH | STSL | STSM |
|---|---|---|---|---|---|---|---|---|---|---|---|---|---|---|---|---|---|---|---|
| HSI1 | 1 | | | | | | | | | | | | | | | | | | |
| HSI2 | .728*** | 1 | | | | | | | | | | | | | | | | | |
| HSI3 | .889*** | .754*** | 1 | | | | | | | | | | | | | | | | |
| HAS | .313*** | .407*** | .315*** | 1 | | | | | | | | | | | | | | | |
| AAIU1 | .077*** | .079*** | .0768*** | .152*** | 1 | | | | | | | | | | | | | | |
| AAIU2 | -.005 | .018 | -.008 | .09*** | .454*** | 1 | | | | | | | | | | | | | |
| AAIU3 | -.01 | .036 | .007 | .067** | .296*** | .313*** | 1 | | | | | | | | | | | | |
| AAIU4 | -.006 | .013 | -.001 | .056** | .295*** | .306*** | .479*** | 1 | | | | | | | | | | | |
| AAIP1 | .058** | .057 | .049* | .105*** | .273*** | .246*** | .095*** | .143*** | 1 | | | | | | | | | | |
| AAIP2 | .054** | .063** | .044* | .129*** | .244*** | .242*** | .072*** | .094*** | .696*** | 1 | | | | | | | | | |
| AAIP3 | .074*** | .064** | .073*** | .075*** | .123*** | .14*** | .068** | .064** | .420*** | .458*** | 1 | | | | | | | | |
| AAIP4 | .039 | .046* | .034 | .025 | .028 | -.002 | .102*** | .138*** | .262*** | .238*** | .289*** | 1 | | | | | | | |
| AAIN1 | -.010 | -.029 | -.020 | -.001 | .132*** | .129*** | .068** | .054** | .248*** | .243*** | .051* | .043* | 1 | | | | | | |
| AAIN2 | -.019 | -.04 | -.034 | .041* | .2*** | .201*** | .074*** | .084*** | .274*** | .259*** | .115*** | .026 | .474*** | 1 | | | | | |
| AAIN3 | -.041^ | -.057** | -.023 | -.068** | .013 | -.016 | .094*** | .126*** | .088*** | .053* | .071*** | .226*** | .291*** | .316*** | 1 | | | | |
| AAIN4 | -.065** | -.055** | -.057** | -.001 | .033 | .052* | .094*** | .112*** | .108*** | .072*** | .069*** | .143*** | .377*** | .381*** | .56*** | 1 | | | |
| STSH | .065** | .033 | .063** | .066** | .247*** | .255*** | .110*** | .094*** | .483*** | .484*** | .261*** | .096*** | .268*** | .289*** | .040*** | .061** | 1 | | |
| STSL | -.042* | -.051* | -.043* | -.0175 | .064** | .073*** | .046* | .102*** | .131*** | .121*** | .071*** | .126*** | .276*** | .305*** | .357*** | .343*** | .124*** | 1 | |
| STSM | .008 | -.011 | .006 | .0164 | .119*** | .186*** | .103*** | .043* | .258*** | .263*** | .207*** | -.029 | .174*** | .203*** | .009*** | .064** | .391*** | .136*** | 1 |

^ $p = .05$, * $p < .05$, ** $p < .01$, *** $p < .001$

**Table 3**. Goodness of fit of measurement and research models

| Model | $\chi^2$/df | CFI | TLI | NFI | RMSEA | SRMR |
|---|---|---|---|---|---|---|
| Measurement model | 581.889/82 = 7.096*** | .961 | .950 | .954 | .051 | .0406 |
| Research model | 1060.473/138 = 7.685*** | .939 | .924 | .931 | .053 | .047 |
| Acceptable criteria (Hu & Bentler, 1998; Kline, 2015) | < 2, $p$ > .05 | > .9 | > .9 | > .9 | < .05 | < .08 |

*** $p < .001$

**Table 4**. Standardized path coefficients in the research model

| | Path | | Standardized coefficient ($\beta$) |
|---|---|---|---|
| Science achievement | ← | Science interest | .349*** |
| AI use | ← | Science interest | -.007 |
| | | Science achievement | .18*** |
| AI positive perception | ← | AI use | .412*** |
| AI negative perception | ← | AI use | .295*** |
| AI positive perception | ↔ | AI negative perception | .297*** |
| STS human and AI | ← | AI positive perception | .225*** |
| | | AI use | .08** |
| | | AI negative perception | .134*** |
| STS life quality | ← | AI positive perception | .472*** |
| | | AI use | .118** |
| | | AI negative perception | .122* |
| STS monopoly | ← | AI positive perception | -.023 |
| | | AI use | -.029 |
| | | AI negative perception | .474*** |
| STS human and AI | ↔ | STS life quality | .251*** |

* $p < .05$, ** $p < .01$, *** $p < .001$

### *Direct, indirect, and total effects of high school science profile*

The indirect and total effects were estimated by bootstrapping (*n* = 2,000) and confidence level inference was made based on bias-corrected confidence level of .95. **Table 5** shows the standardized direct, indirect, and total effects. Straightforwardly, AI used shows highly significant indirect and total effects on the three STS variables (*p* < .01).

Notably, science interest and science achievement showed significantly positive indirect path coefficients on both positive and negative AI perceptions, and the three STS variables (*p* < .05; exceptionally, the indirect effect of science interest on STS life quality is marginally significant with *p* = .05). This first indicates that science achievement may have affected adults' perceptions of AI and views on STS, mediated by their AI use. Second, science interest may also have affected them, mediated by science achievement and AI use.

**Table 5**. Standardized direct, indirect, and total effects

|  | Science interest (indirect) | Science achievement (indirect) | AI use (direct) | (indirect) | (total) |
|---|---|---|---|---|---|
| AI positive perception | .023* | .074** | .412*** | - | .412*** |
| AI negative perception | .016* | .053** | .295*** | - | .295*** |
| STS human and AI | .012* | .038** | .08** | .132** | .213** |
| STS life quality | .019^ | .063** | .118** | .230** | .348** |
| STS monopoly | .006* | .019** | -.029 | .136** | .108** |

^ $p = .05$, * $p < .05$, ** $p < .01$, *** $p < .001$

**Discussions**

*A longitudinal study on science education, AI, and STS*

While there has been a global interest in connecting K-12 science education to the future world, which is hugely influenced by AI and other science and technology, there have not been many studies that empirically show the patterns between the two. This study seems to be one of the earliest reports on the impact of science education on adults' use and perceptions of AI, adding a meaningful piece to our knowledge base.

*Implementation for teaching*

Based on the KEEP II panel data, this study showed that high school science profile predicts adults' AI use, positive and negative perceptions of AI, and views on STS. It is noteworthy as science education has traditionally been endeavoring to foster students' science interest and achievement. This implicates that the principles and know-how that science education community has developed until now can also be effective in preparing students for the future, where AI will be ever more prominent. Echoing previous literature, the results of this study assert it even stronger that science teaching in classroom environments is even more critical in shaping students' future engagement with AI.

*Implementation for research*

Some path coefficients need to be deliberated. It seems that the use of powerful AI services causes both positive and negative perceptions of AI. This can be because the users take advantage of the technology but are also concerned about its long-term consequences. It straight forward that only negative perceptions of AI impacted STS monopoly. However, the significant paths from AI use and both positive and negative perceptions of AI on STS human and AI and STS life quality show that people's standpoint on AI is sophisticated. Probably, even those who have negative perceptions of AI may, counteractively, want to believe in optimistic scenarios. In-depth, qualitative studies are called to further investigate the intricate response from AI users toward AI and STS.

*NARST members' interest*

This study mediates the NARST members' inherent attention in improving science education, and the great driving force of AI that transforms the future-oriented education. This study will attract global attendees of the 2026 NARST Annual Conference, including participants of *Strands 4: Science Teaching — Middle and High School (Grades 5-12)* and *Strands 12: Technology for Teaching, Learning, and Research*, and members of the *RAISE (Research in Artificial Intelligence-Involved Science Education) RIG*. The fruitful discussion at the conference site will enrich the NARST community members' commitment toward the role of science education in the era of AI.

[Version 1 – Last modified: Aug 16, 2025]